\documentclass[twoside]{article}
\usepackage{fleqn,latexsym}

\newcommand{\heads}[2]{\markboth{\protect\small\it #1}{\protect\small\it #2}}
\newcommand{\Acknow}[1]{\subsection*{Acknowledgement} #1}

\newcommand{\bls}[1]{\renewcommand{\baselinestretch}{#1}}

\def\noi{\noindent}

\newcommand{\Title}[1]{\noindent {\Large #1} \\}
\newcommand{\Author}[2]{\noindent{\large\bf #1}\\[2ex]\noindent{\it #2}\\}

\newcommand{\Abstract}[1]{\vskip 2mm \begin{center}
        \parbox{16.4cm}{\small\noi #1} \end{center}\medskip}
\newcommand{\PACS}[1]{\begin{center}{\small PACS: #1}\end{center}}
\newcommand{\foom}[1]{\protect\footnotemark[#1]}

\newcommand{\email}[2]{\footnotetext[#1]{e-mail: #2}}

\newcommand{\Ref}[1]{Ref.\,\cite{#1}}

\newcommand{\sect}[1]{Sec.\,#1}

\def\nq{\hspace{-1em}}
\def\nqq{\hspace{-2em}}
\def\nhq{\hspace{-0.5em}}

\def\cm{\hspace{1cm}}
\def\inch{\hspace{1in}}

\def\eq{Eq.\,}
\def\eqs{Eqs.\,}
\def\beq{\begin{equation}}
\def\eeq{\end{equation}}
\def\bear{\begin{eqnarray}}
\def\al{&\nhq}
\def\lal{&&\nqq {}}               
\def\bearr{\begin{eqnarray} \lal}
\def\ear{\end{eqnarray}}
\def\earn{\nonumber \end{eqnarray}}
\def\dst{\displaystyle}
\def\tst{\textstyle}
\def\nn{\nonumber\\ {}}

\def\nnn{\nonumber\\ \lal }

\def\yy{\\[5pt]}

\def\eql{\al =\al}

\def\e{{\,\rm e}}
\def\d{\partial}

\def\diag{\mathop{\rm diag}\nolimits}

\def\Half{{\dst\frac{1}{2}}}
\def\half{{\tst\frac{1}{2}}}

\def\DAL{\raisebox{-1.6pt}{\large $\Box$}\,}

\newcommand{\arctg}{\arctan}

\newcommand{\sh}{\sinh}

\newcommand{\Arch}{\mathop{\rm Arcosh}\nolimits}
\def\PP{P(\sigma,\rho)}
\def\QQ{Q(\sigma,\rho)}

\bls{1.03}
\heads{ S.V. Chervon and D.Yu. Shabalkin }
{A Plane-Symmetric Gravitational
 Field as a Generalized Nonlinear Sigma Model}

\begin{document}
\twocolumn[

\vspace{-8mm}

\Title{A PLANE-SYMMETRIC GRAVITATIONAL FIELD WITH MATTER\yy
       AS A GENERALIZED NONLINEAR SIGMA MODEL}

\Author{S.V. Chervon\foom 1 and D.Yu. Shabalkin\foom 2}
{Ulyanovsk State University,
42 Leo Tolstoy St., Ulyanovsk 432700, Russia}

\Abstract
{The Einstein equations for a plane-symmetric gravitational field coupled
to an arbitrary nonlinear sigma model (NSM) are shown to be
represented in the form of dynamical equations of a {\it generalized
effective NSM}. The gravitational equations are studied in this case by the
methods of analyzing NSM equations. In the case of a two-component diagonal
NSM exact solutions are found by the functional parameter method.

}

] 
\email{1}{chervon@sv.uven.ru}
\email{2}{sd@sv.uven.ru}

\section{Introduction}

It was proposed in \Ref{chemus} (see also \cite{ch97m}) to consider the
plane-symmetric gravitational field as a four-component
nonlinear sigma model (NSM) defined on a two-dimen\-sional space-time. It
was also shown that some progress can be achieved using the sigma-model
approach.  Nevertheless, it was difficult to obtain a suitable form of
presentation for solutions: the final result of \cite{chemus} involves
integration over an elliptic function. To avoid this problem, the
functional parameter method recently developed \cite{ensm-gc97} and
successfully applied to an effective NSM for plane-symmetric and axially
symmetric gravitational fields \cite{sd98}.  The problem of strong
correspondence between the Einstein equations and the dynamical equations
of a four-component NSM was analyzed in detail in a PhD thesis of one of
the authors \cite{sd98dis}.  Taking into account these results, we intend
in the present paper to generalize the NSM method described above to the
case of  a plane-symmetric space-time filled by chiral fields.  The
two-component NSM as a source of gravity is discussed in detail.

\subsection{Effective NSM for a vacuum plane-symmetric gravitational field}

It was shown in \cite{chemus} that the NSM method may be applied for
studying the Einstein equations for a vacuum plane-symmetric gravitational
field. This connection was introduced with the following construction (see
for details \cite{ch97m}).

The metric
\beq \label{ps}
     ds^2 = A dx^2+2B dx dy+Cdy^2  - D[ dz^2-dt^2]
\eeq
was chosen. The metric coefficients $A,\ B,\ C,\ D$ depend on $z$ and $t$
only.  Following the ideas of the effective NSM method, let us
introduce an NSM with the chiral fields
$\varphi^1=\psi$, $\varphi^2=\theta$, $\varphi^3=\chi$, $\varphi^4=\phi$,
defined on the space-time
$g^{(e)}$: $ds^2_{(e)}=dz^2-dt^2$, and taking their values in the
target space
\beq	\label{chispace}
h_{AB}^{(e)}={e^\psi}\left(\matrix {-1&0&0&-1\cr
                                 0&1&0&0\cr
                                 0&0&\sh^2\theta&0\cr
                                 -1&0&0&0\cr}\right).
\eeq
The dynamic equations derived from the action
\beq\label{Snsm}
{\cal S}= \int\limits_{\cal M} \sqrt{|g^{(e)}|}\ d^4x \biggl\{\frac{1}{2}
 h^{(e)}_{IK}(\varphi) \varphi^I_{,i} \varphi^K_{,k} g^{(e)ik} \biggr\}
\eeq
have the form
\bearr     \label{effNSM}
	\DAL\e^\psi  = 0,\nnn
\DAL\theta+(\psi_z\theta_z-\psi_t\theta_t)- \half (\chi^2_z-\chi^2_t)
  	\sinh 2\theta = 0, \nnn
\DAL\chi+2(\theta_z\chi_z-\theta_t\chi_t)\coth\theta
 	+(\psi_z\chi_z-\psi_t\chi_t) = 0,\nnn
\DAL\phi+ \half (\psi^2_z-\psi^2_t) - \half
		(\chi^2_z-\chi^2_t)\sinh^2\theta \nnn  \inch
	-\half (\theta^2_z-\theta^2_t) = 0, \\ \lal
				\DAL\equiv\d_{zz}-\d_{tt}.
\earn

The Einstein vacuum equations $R_{ik}=0$ for the metric (1)
are reproduced from the dynamic equations (4) if the metric coefficients are
connected with the chiral fields in the following way:
\bear\label{CFMC}
A \eql -\e^\psi({\rm cos}\chi \sinh\theta+\cosh\theta),\nn
B \eql \e^\psi \sin\chi \sinh\theta,\nn
C \eql \e^\psi (\cos\chi \sinh\theta-\cosh\theta),\nn
D \eql \e^\phi.
\ear

This direct connection between the gravitational field and the ef\-fec\-tive
chiral fields allows one to transfer the analysis from the Einstein vacuum
equations for (\ref{ps}) to the effective NSM equations (\ref{effNSM}).

The effective NSM equations (\ref{effNSM}) were studied
by means of the isometric ansatz method and the functional parameter method
\cite{chemus,shch-97,ensm-gc97,sd98}.
The functional parameter method provides a wider family of solutions
and should be presented here for further constructions.
Solutions of (\ref{effNSM}) will be sought in the form
\beq\label{Sol0}
   \theta=\theta(\xi),\quad \chi=\chi(\xi),
   \quad \phi=\phi(\xi), \quad   \psi=\ln \xi
\eeq
where $\xi=\xi(z,t)$ is a functional parameter satisfying
\beq\label{Solxi}
  	\DAL\xi=0,
\eeq
according to the first equation (\ref{effNSM}).

Substitution of (\ref{Sol0}) into the other equations (\ref{effNSM})
leads to the set of ordinary differential equations
\bear  \label{ODU}
  \ddot\theta+\frac{1}{\xi}\dot\theta
  		-\frac{1}{2}\dot\chi^2\sinh 2\theta \eql 0,\nn
  \ddot\chi+\frac{1}{\xi}\dot\chi+{2}\dot\chi\dot\theta\coth\theta \eql 0,\nn
  \ddot\phi+\frac{1}{2\xi^2}-\frac{1}{2}\dot\theta^2
  				-\frac{1}{4}\dot\chi^2\sinh^2\theta \eql 0
\ear
for the functions $\theta$, $\chi$ and $\phi$.
The parameter $\xi$ is treated here as an independent variable.
Here and henceforth
\[
     \dot{\phi_a}=\frac{d}{d\xi}\phi_a, \qquad
     \ddot{\phi_a}=\frac{d^2}{d\xi^2}\phi_a,\ldots;\qquad  a=1,2,3,4.
\]
\eqs (\ref{ODU}) are exactly integrated and the solutions
may be written as
\bear\label{FS}
  \psi \eql \ln\xi,	\cm 	\DAL\xi(z,t)=0;\\
\theta \eql \Arch\left[{k\over 2}\left(\left[\xi/\xi_0\right]^a+
 	        		\left[\xi/\xi_0\right]^{-a}\right)\right];\\
\chi_{\pm}  \eql \arctg\chi_0\pm \arctg\left[\dst\left|
	\frac{a}{c}\right|\dst\frac{\left[\xi/\xi_0\right]^{2a}+1}
		{\left[\xi/\xi_0\right]^{2a}-1}\right];\\
\phi	\eql  \phi_0+\phi_1\xi+\frac{a^2-1}{2}\ln \xi, \label{FSlast}
\ear
where $k=\sqrt{(a^2+c^2)/a^2}$; $a$ and $c$ are arbitrary constants. The
metric is obtained by substituting (\ref{FS})--(\ref{FSlast})
into (\ref{CFMC}).

The case of vacuum axially and plane-symmetric space-times can be thus
investigated by means of the functional parameter method \cite{sd98}.

An evident generalization of the effective NSM method is to
plane-symmetric gravitational field filled by matter
represented by physical fields, e.g., chiral ones.
It is preferred to study such systems by NSM analysis only.  In other
words, construction of a generalized effective NSM, including gravity
as well as a physical NSM, is desired.  In \sect 2 we consider
the Einstein equations $R_{ik}=T_{ik}-{1\over2} g_{ik} T$ for the energy
momentum tensor $T_{ik}$ corresponding to an arbitrary $M$-component NSM.
The possibility of representing both the Einstein equations and the NSM
dynamic equations in terms of a {\it generalized} effective NSM will be
analyzed.

A complete analysis of a plane-symmetric gra\-vi\-ta\-tio\-nal space filled
with a 2-component nonlinear sigma model is carried out in \sect 3.
Exact solutions are presented there.

\section{Generalization of the effective NSM method}\label{equat}

Consider a space-time (\ref{ps}) created by an NSM described by the chiral
fields $\Phi^1(z,t),\ldots,\Phi^M(z,t)$, taking their values on  some
$h_{IK}(\Phi)$. Let us accept the following rules for the indices:
$A,B =1,\ldots, 4$ and $I,J,K =1, \ldots, M$.

The system under consideration will be governed by the Einstein equations
\beq\label{EE}
		R_{ik}=T_{ik}-{1\over2} g_{ik} T
\eeq
and the NSM field equations
\bear\label{ENSM}
\lal h_{JK}\DAL \Phi^K+{1\over \alpha}(\alpha_t\Phi_t-\alpha_z\Phi_z)\nnn
\cm + (h_{IK,J}-\Half h_{JK,I})(\Phi^I_t\Phi^K_t-\Phi^I_z\Phi^K_z)=0,\nnn
\alpha=\sqrt{AC-B^2}.
\ear

To construct a generalized effective NSM it is necessary:

\begin{itemize}
\item
to represent the gravity equations (\ref{EE}) as dynamic equations of
a certain NSM, i.e. to define an effective
space-time ($ds^2_{(e)}$) containing effective chiral fields,
the effective chiral space $h^{(e)}_{IK}$
and a connection between the metric coefficients and the chiral fields
similar to the vacuum case;

\item
to add the physical NSM (\ref{ENSM}) into this model.
\end{itemize}

Having satisfied these requirements, one can construct a generalized chiral
space in the form
\[
	H_{LN}=H_{LN}(h^{(e)}_{AB},h_{IK}),\quad
		(L,N,P = 1\ldots M+4)
\]
with the fields $\Psi=(\varphi^1,\ldots,\varphi^4,\Phi^1,\ldots,\Phi^M)$
defined in the space-time with $g^{(e)}_{ik}$ and the corresponding
Lagrangian
\beq\label{gLagr}
{\cal L}={1\over 2} \sqrt{|g^{(e)}|}g^{(e)\,ik}H_{LN} \Psi^L_i \Psi^N_k.
\eeq
The equations due to (\ref{gLagr})
\bearr\label{GeNSM}
 \frac{1}{\sqrt{|g^{(e)}|}} (H_{NL}\Psi^L_k g^{(e)\,ik})_i-
 		{1 \over 2} H_{NL,P}\Phi^L_i\Phi^P_k g^{(e)\,ik}=0 \nnn
\ear
should unify (\ref{EE}) and (\ref{ENSM}).

Let us show now that the Einstein equations are contained in the chiral
model equations.  \eqs (\ref{EE}) with the chosen energy-momentium tensor
may be written as $R_{ik}=h_{IK}\Phi^I_i\Phi^K_k$. For $i,k=1,2$ they are
the same as in the vacuum case $R_{ik}=0$ and therefore may be represented as
(\ref{GeNSM}).  They follow from (\ref{GeNSM}) with $N=2,3,4$. The equation
with $N=1$ will be satisfied by the Einstein equations with $i,k=3,4$.

The original set of equations (\ref{EE}), (\ref{ENSM}) in terms of the
NSM will be written in the following way:
\bear					\label{GENSM1}
		\DAL\e^\psi \eql 0,\nn
\DAL\theta+(\psi_z\theta_z-\psi_t\theta_t)-{1\over2}(\chi^2_z-\chi^2_t)
  		\sinh 2\theta \eql 0,
					\nn
\DAL\chi+2(\theta_z\chi_z-\theta_t\chi_t)\coth\theta
 		+(\psi_z\chi_z-\psi_t\chi_t) \eql 0,
					\nn
\DAL\phi+{1\over2}(\psi^2_z-\psi^2_t)
  		-{1\over2}(\chi^2_z-\chi^2_t)\sinh^2\theta  \lal \nn
- {1\over2}(\theta^2_z-\theta^2_t)
		-h_{AB}(\Phi^A_t\Phi^B_t-\Phi^A_z\Phi^B_z) \eql 0,\nn
							 \nqq \\
\label{GENSM2}
	h_{JK}\DAL\Phi^K+(\psi_t\Phi_t-\psi_z\Phi_z)   \cm \lal\nn
+ (h_{IK,J}-{1 \over 2} h_{JK,I})(\Phi^I_t\Phi^K_t-\Phi^I_z\Phi^K_z)
							\eql 0,       \nn
J=1\ldots M.       \inch
\ear

All the equations are NSM dynamic equations. It is now necessary to unify
them in a generalized effective model corresponding to (\ref{gLagr}) by
choosing the form of $H_{LN}$ and $g^{(e)}_{ik}$.  \eqs (\ref{ENSM})
give the following possible form of the generalized chiral space metric:
\beq\label{GeChSp}
	H_{LN}=\left(\matrix{h^{(e)}_{AB}&0\cr
        	                        0&H_{IK}\cr}\right).
\eeq
\eq (\ref{GENSM1}.4), containing the term
$
	h_{IK}(\Phi^I_t\Phi^K_t-\Phi^I_z\Phi^K_z),
$
can arise from the generalized effective NSM with the chiral space metric
(\ref{GeChSp}) if
\beq\label{HAB}
	H_{IK}=-2 \e^\psi h_{IK}.
\eeq
Note that \eqs (\ref{GENSM1}) are constructed under the assumption
$g^{(e)} = \diag (1,-1)$ but with the metric $g_{ik}$
for the physical NSM [\eqs (\ref{GENSM2}) have the form (\ref{ps})].
However, the NSM equations with the physical chiral
fields $\Phi^I$, defined on $g^{(e)}=\diag (1,-1)$ with the chiral space
in the form (\ref{HAB}), will coincide with the physical NSM
equations (\ref{ENSM}).

In this way the NSM formed by the chiral fields
\beq\label{GEF}
	\Psi^L=(\varphi^1,\ldots\varphi^4,\Phi^1,\ldots,\Phi^M),
\eeq
defined on the space-time with the metric
\beq\label{GEST}
	ds_{(e)}^2=dz^2-dt^2
\eeq
with the chiral space corresponding to
\bearr\label{GECS}
dS^2 = \e^\psi(-d\psi^2+d\theta^2+\sinh^2\theta d\chi^2 \nnn \cm
		-2 d\psi d\phi
			-2 h_{IK}d \Phi^I d\Phi^K)
\ear
may be treated as a generalized effective nonlinear sigma model of
the plane-symmetric space-time (\ref{ps})
filled with a physical NSM of general form. The dynamic equations
(\ref{GENSM1}), (\ref{GENSM2}) of the generalized effective NSM
(\ref{GEF}-\ref{GECS}) contain the Einstein equation (\ref{EE}) for
the metric (\ref{ps}) with the physical NSM (\ref{EE}) and the dynamic
equations of physical chiral fields (\ref{ENSM}).

\section{Exact solutions}\label{es}

Consider a special case of a physical NSM, namely, a two-component NSM,
represented by $\Phi^1=\sigma(z,t)$, $\Phi^2=\rho(z,t)$ and
\[
h_{AB}=\left(\matrix{P(\sigma,\rho)&0\cr
                       0&Q(\sigma,\rho)}\right).
\]

The effective fields $\psi,\theta,\chi$ and therefore, according to
(\ref{CFMC}), the metric coefficients $A,\ B,\ C$ are surely the same as
in empty space \cite{ch97m}. Solutions of these equations have been obtained
by the method of functional parameter (\ref{FS}).  The fourth equation,
describing $\phi=\ln D$, will have the form
\bearr						\label{SEE4}
	\DAL\phi+{1\over2}(\psi^2_z-\psi^2_t)
		-{1\over2}(\chi^2_z-\chi^2_t)\sinh^2\theta \nnn
\ -{1\over2}(\theta^2_z-\theta^2_t) + P(\sigma_z^2-\sigma_t^2)+
		Q(\rho_z^2-\rho_t^2)=0.
\ear
Under the assumption (\ref{FS}) this equation may be written in the form
\beq\label{SEE4pr}
\ddot\phi+\frac{1}{2 \xi^2}(a^2-1)+ \PP\dot\sigma^2+\QQ\dot\rho^2=0.
\eeq

This equation may be analyzed either by direct integration
by methods applicable to NSM or by the modified
method of solutions generation from a vacuum seed solution \cite{ch97m}
by putting $\phi=\phi_{(v)}+\phi_{(m)}$.
Here $\phi_{(v)}$ is a solution to the vacuum part of \eq (\ref{SEE4}),
corresponding to empty space (\ref{FS}).
The term $\phi_{(m)}$ is the matter correction and in the present case
satisfies the equation
\beq\label{MEE4}      \nq\
\DAL\phi_{(m)}+\PP(\sigma_z^2-\sigma_t^2)+\QQ(\rho_z^2-\rho_t^2)=0.
\eeq

The fields $\sigma$ and $\rho$ are solutions of
(\ref{GENSM2}) written for our NSM as follows:
\bearr\label{SNSM}
		P\DAL\sigma-\half\ Q_{\sigma}(\rho_z^2-\rho_t^2)+
			\half P_{\sigma}(\sigma_z^2-\sigma_t^2)\nnn
\quad + P_{\rho}(\sigma_z\rho_z-\sigma_t\rho_t)+
			P(\psi_z\sigma_z-\psi_t\sigma_t)=0,\nnn
	Q\DAL\sigma+\half\ Q_{\rho}(\rho_z^2-\rho_t^2)-
			\half P_{\rho}(\sigma_z^2-\sigma_t^2)\nnn
\quad + Q_{\sigma}(\sigma_z\rho_z-\sigma_t\rho_t)+
				Q(\psi_z\rho_z-\psi_t\rho_t)=0.
\ear

To solve these equations, let us apply the
functional parameter method with
$\psi=\ln\xi,\,\sigma=\sigma(\xi),\,\rho=\rho(\xi)$.
The set of ordinary differential equations set may be written as
\bearr\label{FE}\nq\
    \ddot\sigma
	-\Half {Q_\sigma\over P}\dot\rho^2+\Half(\ln P)_\sigma\dot\sigma^2+
	(\ln P)_\rho \dot\sigma\dot\rho+{\dot\sigma\over\xi}=0,
\nnn\nq\
\ddot\rho-\Half {P_\rho\over Q}\dot\sigma^2+\Half(\ln Q)_\rho\dot\rho^2+
	(\ln Q)_\sigma \dot\sigma\dot\rho+{\dot\rho\over\xi}=0.
\ear
These equations can hardly be exactly integrated for arbitrary $\PP$ and
$\QQ$. Here we discuss some special cases.

\medskip\noi
{\bf 1.} The simplest case, $\PP=\QQ=1$:
\bear
	\sigma(\xi)\eql\sigma_0+\sigma_1\ln\xi,\nn
	  \rho(\xi)\eql\rho_0+\rho_1\ln\xi,\\
\phi\eql \Half (a^2-1)+(\sigma_1^2+\rho_1^2)
			\ln \xi+\phi_1 \xi+\phi_0.	\label{simple}
\ear

\noi {\bf 2.} $\PP=p(\sigma),\,\QQ=q(\sigma)$:
\bearr
	\xi \frac{d \rho}{d \xi}=\frac{c}{q(\sigma)}, \nnn\nq
\biggl(\xi^2\frac{d^2}{d\xi^2}+\xi \frac{d}{d \xi}\biggr)\sigma
    + \Half \xi^2 \frac{d \ln p(\sigma)}{d\xi}\frac{d\sigma}{d\xi}-
			\frac{c^2}{2 p q^2}\frac{d q}{d \sigma}=0.
\earn

\noi
{\bf 2.1.} Rotation surface, $\PP=1,\, \QQ=q(\sigma)$:
\[
\xi \frac{d \rho}{d \xi}=\frac{c}{q(\sigma)},\cm
	\frac{d\xi}{\xi}=\sqrt{k-\frac{c^2}{q(\sigma)}}d\sigma.
\]

\noi
{\bf 3.} $\PP=p(\sigma),\,\QQ=q(\rho)$:
\beq\label{Vac}
	\sqrt{p(\sigma)}d\sigma=\sigma_0\frac{d\xi}{\xi} ,
		\qquad \sqrt{q(\rho)}d\rho=\rho_0\frac{d\xi}{\xi}.
\eeq
This case is of particular interest. In the case (\ref{Vac}) matter
appears in the form
\[
	\ddot\phi_{(m)}=\frac{1}{\xi^2}(\sigma_0^2+\rho_0^2),
\]
which leads to
\[
\phi_{(m)}=\phi_m^{(0)}+\phi_m^{(1)}\xi+(\sigma_0^2+\rho_0^2)\ln \xi.
\]
The above expression is similar to the vacuum case,
\[
	\phi_{(v)}=\phi_0+\phi_1\xi+\frac{a^2-1}{2}\ln \xi.
\]
Then the solution is
\bearr\label{Phi}
\phi = \biggl(\frac{a^2-1}{2}+\sigma_0^2+\rho_0^2\biggr)\ln \xi \nnn
	\cm +(\phi_m^{(1)}+\phi^{(1)})\xi + (\phi_m^{(0)}+\phi^{(0)}).
\ear
This expression for the field $\phi$ coincides with the solution
(\ref{simple}).  This means that cases under consideration is a sort of
gauge transformation of chiral space.  Solution for $\phi$ (for $D$) in the
presence of matter, by a proper choice of the constants (under special
initial conditions) may lead to the same results as in the vacuum case.
The chiral space corresponding to $h_{AB}$ may be then
represented as a non-deformed plane which may be curved in any way.

\section{Conclusions}

In this paper the effective NSM method has been extended to
plane-symmetric space-times filled by chiral fields.
The functional parameter method was shown to be applicable for generating
exact solutions for such systems.  It opens possibilities of
construction and study of exact cosmological solutions in the spirit
of Belinskii \cite{BelVAWE}, with sources represented by nonlinear sigma
models. A family of such solutions was pointed out in \cite{ch97m}, and any
classes of solution may be constructed in the future by the method suggested.

\Acknow
{This work has been carried out under the aegis of the State
Research Programme ``Astronomy. Fundamental Space Research'', Section
``Cosmomicrophysics,'' and with partial financial support from the Russian
Basic Research Foundation (Grant No. 98-02-18040) and the Centre of
Cosmoparticle Physics ``Cosmion''. The authors thank V.M. Zhuravlev
for useful discussions.  }


\small
 \begin{thebibliography}{99}

\bibitem{matmis67}
R.A. Matzner and C.W. Misner,
{\it Phys. Rev. } {\bf 154}, 1229 (1967).

\bibitem{chemus}
Chervon S.V., Muslimov A.G.,
{\it Phys.Lett.} {\bf A142}, 1989, P.14. 

\bibitem{ch97m}
S.V. Chervon,
``Non-Linear Fields in the Theory of Gravitation and Cosmology'',
Middle-Volga Scientific Centre, Ulyanovsk State University, 1997.

\bibitem{ensm-gc97}
S.V. Chervon, V.M. Zhuravlev and D.Yu. Shabalkin,
{\it Grav. \& Cosmol.\/} {\bf 3} (1997) , No. 4(12), 312-316.

\bibitem{shch-97}
D.Yu. Shabalkin and S.V.Chervon,
{\it Izvestiya Vuzov. Fizika} 1998, No 6 P.114 (in Russian)

\bibitem{sd98}
Shabalkin D.Yu.
{\it Grav. \& Cosmol.\/} {\bf 4}, 4 (16), 302-307 (1998).

\bibitem{sd98dis}
Shabakin D.Yu.,
``Effective Nonlinear Sigma Models in Gravitation and Cosmology'',
Ulyanovsk, 1998, PhD thesis (in Russian).

\bibitem{BelVAWE} V.A.~Belinskii,
{\it Zh. Eksp. Teor. Fiz.} {\bf 77}, 1239--1254 (1979).

\end{thebibliography}
\end{document}